\documentclass[11pt,twoside]{article}
\usepackage{epsfig}
\usepackage{here}
\usepackage{amsmath}
 
\begin{document}
 

\begin{center}{\bf\Large Feasibility study of the 
                   quasi-free creation of the $\eta^{\prime}$ 
                   meson in the reaction $pn\rightarrow pn\eta^{\prime}$}
\end{center}
\vspace{0.5cm}
\begin{center}
   P. Moskal$^{a,b}$ \\[0.1cm]
   {\small \em 
         $^a$Institute of Physics, Jagellonian University, Krak\'ow, Poland \\
         $^b$Institut f\"ur Kernphysik, Forschungszentrum-J\"ulich, Germany
   }
\end{center}
\vspace{0.5cm}
\begin{center}
 \parbox{0.9\textwidth}{
  \small{
    {\bf Abstract:}\
      The  feasibility of an investigation of the $pn\rightarrow pn\eta^{\prime}$
      reaction by means of the COSY-11 internal target facility is discussed. 
      Appraisals
      are based on the assumption of the quasi-free reactions 
      of beam protons, circulating in the cooler synchrotron COSY,
      with neutrons from a windowless deuteron cluster target.
 }
 }
\end{center}

\vspace{0.5cm}
\section{Introduction}\vspace{-1cm}\hspace{4cm}\footnote{
         This part of the talk presents some aspects 
         of the motivation from the 
         COSY Proposal$\#$100~\cite{proposal100}}\\
       
\begin{figure}[H]
        \vspace{-0.4cm}
        \parbox{0.3\textwidth}{\epsfig{file=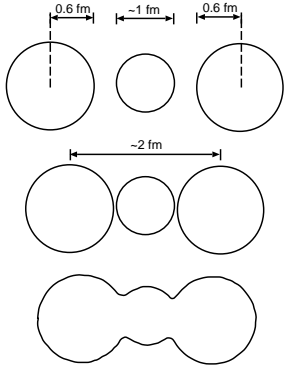,width=0.3\textwidth,angle=0}}
        \hfill
        \parbox{0.5\textwidth}{\caption{\small A cartoon illustrating in naive geometrical terms that
                        for $r < 2r_{N} + 2r_{M}$ meson exchange is unlikely to be
                        appropriate to the description of the internucleon potential.
                        Figure and caption are taken from reference~\cite{maltmanisgur}.
        \label{cartoon}
        }}
      \end{figure}

      In 1984 Maltman and Isgur~\cite{maltmanisgur}
      have argued on the basis of simple geometrical 
      considerations (see Figure~\ref{cartoon}) 
      that  at the distances smaller than 2~fm 
      the internucleon potential should begin to be free
      of meson exchange effects and 
      may  be \hfill dominated \hfill 
      by the \hfill residual \hfill colour \hfill forces. \hfill However, as
      \hspace{-0.6cm}pointed out by Nakayama~\cite{nakayamaproc},
      the transition region from the
      hadronic to constituent quark degrees of freedom
      does not have a well defined boundary, and at present
      both approaches should be evaluated in order to test 
      their relevance in the description of 
      close-to-threshold  meson production
      in the collision of nucleons.
      The authors of
      reference~\cite{kleefeld_ss} have shown, that 
      for example the $K^{+}$ meson production via the $pp\rightarrow pK^{+}\Lambda$
      reaction far from its production threshold 
      can well be described in terms of  either
      the meson-exchange mechanism or the two-gluon exchange model.
      Thus, a determination of the relevant 
      degrees of freedom for the description of the nucleon-nucleon
      interaction, especially in case when nucleons are very close 
      together, remains one of the key issues in the hadronic physics~\cite{capstick}.

\begin{figure}[H]
        \vspace{-0.4cm}
        \centerline{\epsfig{file=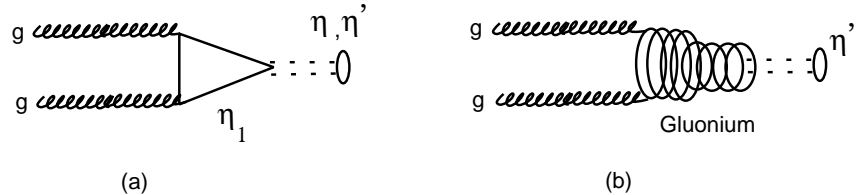,width=0.9\textwidth,angle=0}}
        \vspace{-0.5cm}
        \caption{\small Coupling of $\eta$ and $\eta^{\prime}$ to two gluons through
          \ \ \ (a) quark and antiquark triangle loop  and \ \ (b) gluonic admixture.
          The figure is taken from reference~\protect\cite{kou99}.
        \label{triangle}
        }
      \end{figure}

      Close-to-threshold production of $\eta$ and $\eta^{\prime}$ mesons
      in the nucleon-nucleon interaction requires a large momentum 
      transfer between the nucleons and hence can occur only at distances
      smaller than $\sim$0.3~fm.  This suggests that the quark-gluon 
      degrees of freedom may indeed
      play a significant role in the production dynamics of these mesons,
      and especially that the $\eta^{\prime}$  meson can be created directly 
      from the glue which is excited in the interaction region of the colliding 
      nucleons~\cite{basswetzel,kolacosynews}.
      A possibly large glue content of the $\eta^{\prime}$ and the dominant
      flavour-singlet combination of its quark wave function
      may cause that the
      dynamics of  its production process in the nucleon-nucleon
      collisions is significantly different from that responsible for the 
      production of the $\eta$ meson  
      (see Figure~\ref{triangle}). 
      \begin{figure}[H]
        \vspace{-0.7cm}
        \centerline{\epsfig{file=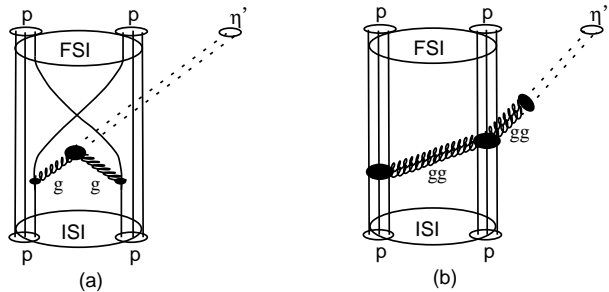,height=4.0cm,angle=0}}
        \caption{ \small 
           Diagrams depicting
             possible quark-gluon dynamics of the reaction $pp \rightarrow pp\eta^{\prime}$.
          (a)--- production via a fusion of gluons~\protect\cite{kolacosynews}
                 with rearrangement of quarks.
          (b)--- production via a rescattering of a
                 ``low energy pomeron''~\protect\cite{basspriv}.
        \label{graf_glue}
        }
      \end{figure}
      Figure~\ref{graf_glue} depicts possible short-range mechanisms 
      which may lead to the creation of the $\eta^{\prime}$ meson
       via a fusion of gluons emitted from the exchanged
      quarks of the colliding protons~\cite{kolacosynews} or via an
      exchange of a colour-singlet object made up from glue, which then
      rescatters and converts into $\eta^{\prime}$\cite{basspriv}.
      The hadronization of gluons to  the $\eta^{\prime}$ meson
      may proceed directly via its gluonic component 
      or through its overwhelming flavour-singlet admixture~$\eta_{1}$~(see Fig.~\ref{triangle}). 
      In contrary to the significant meson exchange mechanisms and the fusion of gluons
      of graph~\ref{graf_glue}a, the creation through the colour-singlet object proposed
      by S.~D.~Bass~\cite{basspriv} (graph~\ref{graf_glue}b) is isospin independent, and hence should
      lead to the same production yield\footnote{after 
                       correcting for the final and initial state interaction between 
                       participating baryons.}
      of the $\eta^{\prime}$ meson 
      in both reactions: $pp\rightarrow pp\eta^{\prime}$ and 
      $pn\rightarrow pn\eta^{\prime}$ 
      because gluons do  not
      distinguish between flavours.
      This property should allow to test the 
      relevance of 
      a short range gluonic term~\cite{bassproc}
      by the experimental determination of the cross section ratio
      $R_{\eta^{\prime}}=\sigma(pn\rightarrow pn\eta^{\prime})
                        /\sigma(pp\rightarrow pp\eta^{\prime})$,
      which in that case 
      should be close to unity after correcting for the final and 
      initial state  interaction.
      The other extreme scenario --~assuming the dominance of the 
      isovector meson exchange mechanism~-- should result in the value of $R_{\eta^{\prime}}$
      close to 6.5 as was already 
      established in the case of the  $\eta$ meson~\cite{calen_pneta}.
        
      The total cross section for the $pp\rightarrow pp\eta^{\prime}$ reaction
      has already been measured close to the kinematical threshold
      by the COSY-11~\cite{moskalprl,moskalpl},
      SPES-III~\cite{hiboupl} and DISTO~\cite{disto_etap} collaborations. 
      However, data on 
      the near threshold production of the $\eta^{\prime}$ meson 
      in proton-neutron collisions do not exist.
      Thus as a first step towards
      the determination of the value of $R_{\eta^{\prime}}$ the
      feasibility of the measurement of the $pn\rightarrow pn\eta^{\prime}$
      reaction by means of the COSY-11 facility was studied by the Monte-Carlo
      method and is discussed in the following section.

\section{Experimental method: Quasi-free production }
\vspace{-1cm}\hspace{12.2cm}\footnote{
        Results presented in this section constitute 
        an extended experimental
        part of the COSY Proposal$\#$100~\cite{proposal100}
       }\\

\begin{figure}[H]
       \vspace{-1.0cm}
       \parbox{0.5\textwidth}{\epsfig{file=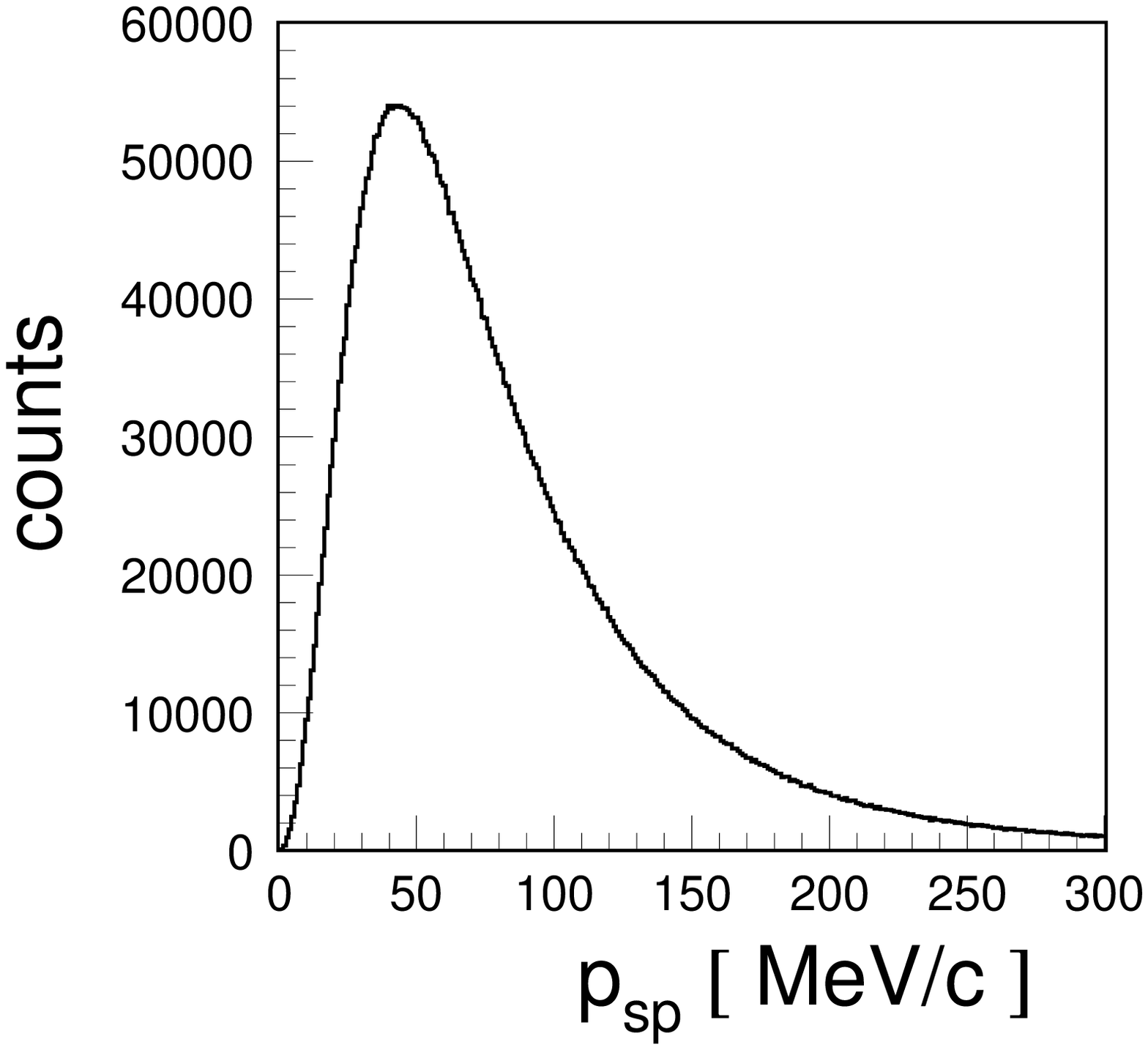,width=0.5\textwidth,angle=0}}
       \parbox{0.5\textwidth}{\epsfig{file=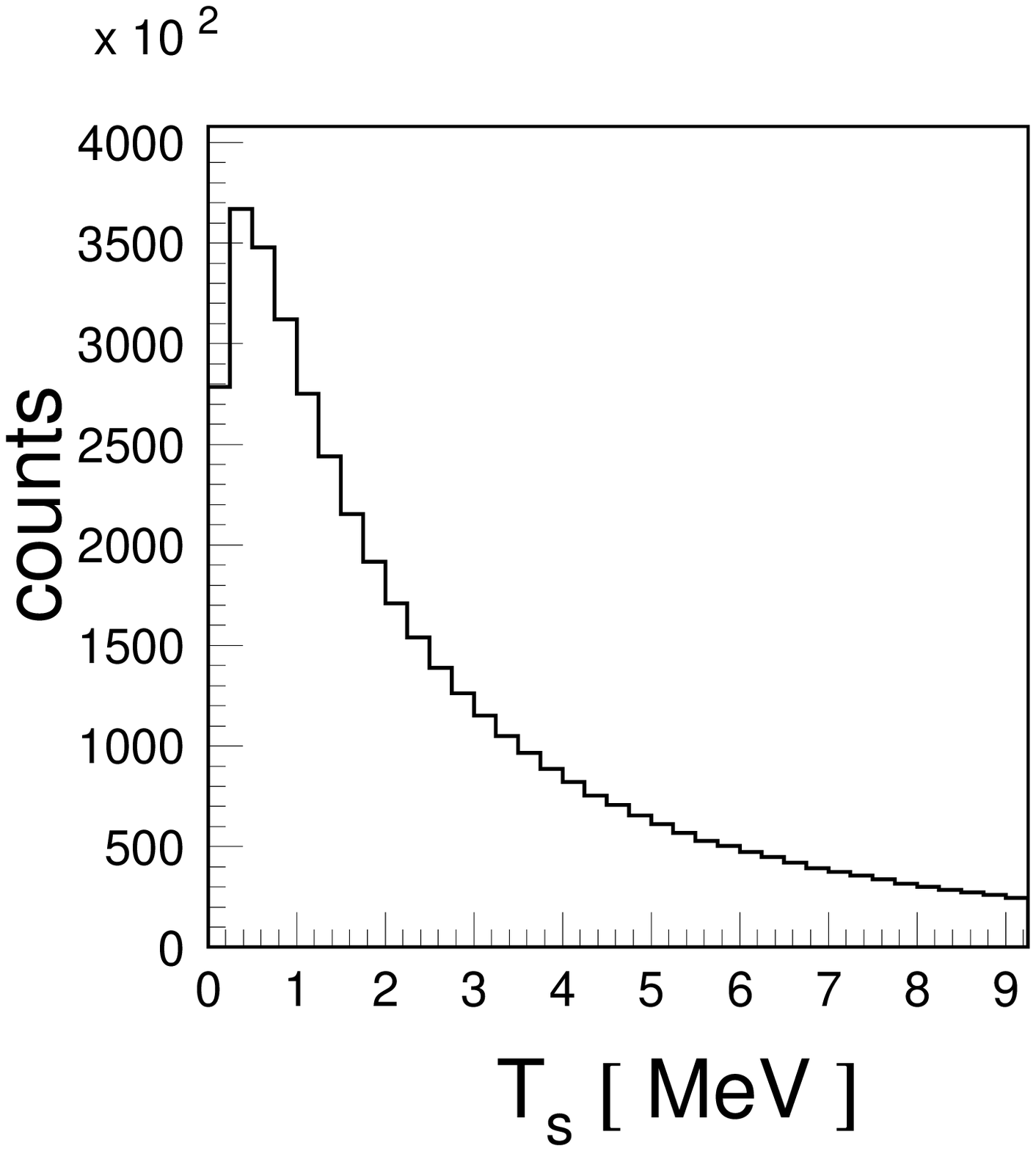,width=0.5\textwidth,angle=0}}

       \vspace*{-0.9cm}

       \parbox{0.45\textwidth}{\mbox{} }\hfill
       \parbox{0.51\textwidth}{\large a) }\hfill
       \parbox{0.04\textwidth}{\large b) }

   \caption{ \small
           (a) Momentum and (b) kinetic energy
           distribution of the nucleons in the deuteron, 
           generated according to an analytical parametrization of the 
           deuteron wave function~\protect\cite{lacombe81,stina}
           calculated from the PARIS potential~\protect\cite{lacombe80}.
           \label{impuls_and_spec_mom}
         }
\end{figure}

\begin{figure}[h]
       \vspace{-1.0cm}
       \parbox{0.5\textwidth}{\epsfig{file=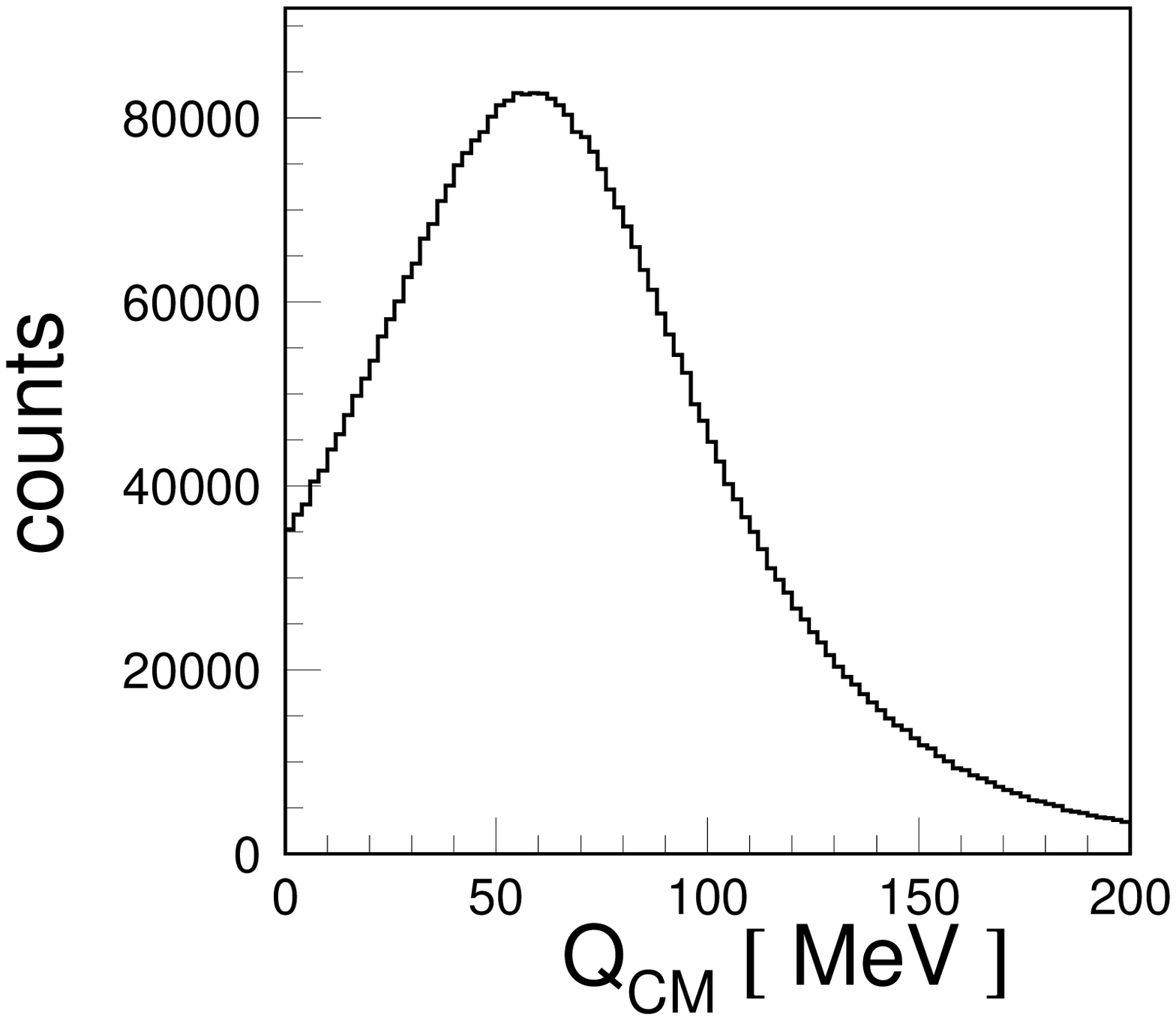,width=0.5\textwidth,angle=0}}
       \parbox{0.5\textwidth}{\epsfig{file=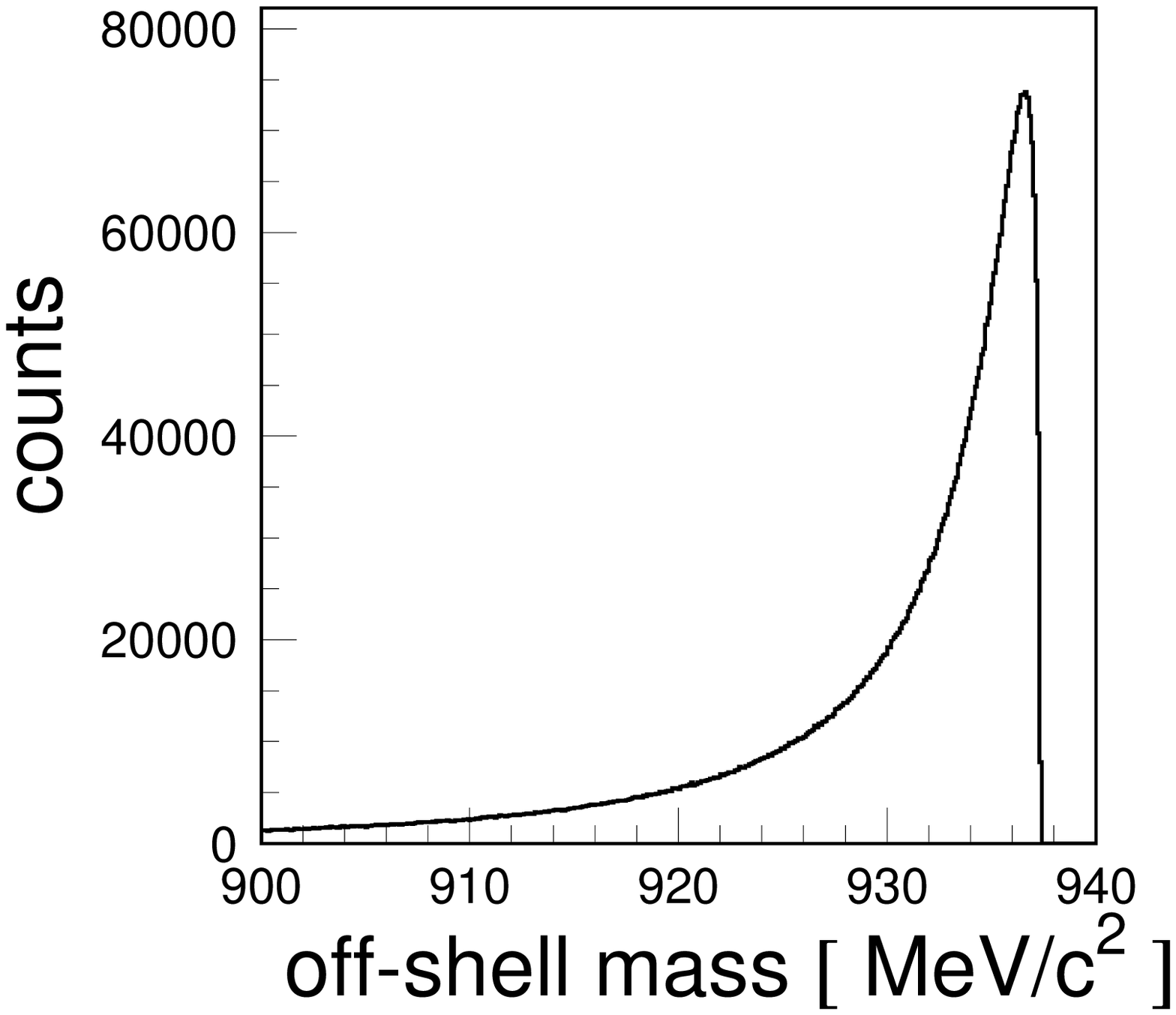,width=0.5\textwidth,angle=0}}

       \vspace*{-0.9cm}

       \parbox{0.45\textwidth}{\mbox{} }\hfill
       \parbox{0.51\textwidth}{\large a) }\hfill
       \parbox{0.04\textwidth}{\large b) }

   \caption{ \small
         (a) Distribution of the excess energy $Q_{CM}$ 
           for the $pn\eta^{\prime}$ system originating
           from the reaction $pd \rightarrow p_{sp} pn\eta^{\prime}$
           calculated with a proton beam momentum of 3.350~GeV/c,
           and the neutron momentum smeared out according to the 
           Fermi distribution shown in Figure~\protect\ref{impuls_and_spec_mom}b.
            The beam momentum of 3.350~GeV/c corresponds
           to an excess energy of $Q_{CM} = 44.7$~MeV for the 
           free $pn\rightarrow pn\eta^{\prime}$ reaction. \protect\\
         (b) Spectrum of the  off-shell mass of the interacting neutron,
           as calculated under the assumption of the impulse approximation. 
           \label{Q_and_mass_off}
         }
\end{figure}

  In order to measure the $pn\rightarrow pn\eta^{\prime}$ reaction
by means of a proton beam  it is necessary to use a nuclear target,
since a pure neutron target does not exist. 
Similar to
investigations of the $\eta$ meson production
in the $pn\rightarrow pn\eta$ reaction~\cite{chiavassa94,calenpdeta,calen_pneta},
a deuteron will be considered as  source of neutrons. 
Due to the small binding energy of the deuteron ($E_{B}~=$~2.2~MeV),
the neutron struck by the proton incoming with 2540~MeV kinetic energy 
may approximately be  treated as  being a free particle
in the sense that the matrix element for quasi-free meson production
on a bound neutron is identical to that for the free $pn\rightarrow pn\eta$
reaction.  Measurements performed at CELSIUS~\cite{calenpdeta,calen_pneta}
and TRIUMF~\cite{duncan,hahn} have proven that the offshellness of the reacting 
neutron can be neglected and that  the spectator proton influences the interaction
only in terms of the associated Fermi motion~\cite{duncan}.
In this approximation the proton from the deuteron
is considered as a spectator
which does not interact with the bombarding proton, but rather
escapes untouched and hits the detectors  carrying the Fermi momentum
possessed at the moment of the collision.
  The momentum spectrum of the 
nucleons in the deuteron is shown in Figure~\ref{impuls_and_spec_mom}a. 
Since the 
neutron bound inside a deuteron is not at rest, and its momentum may change from event to event,
the excess energy in the quasi-free proton-neutron reaction will also
vary from event to event. This enables to scan a large range of 
excess energies with a constant proton beam momentum, but simultaneously
requires the determination of this energy  for each registered event,
which can be done only if the neutron momentum vector is known. 
This may be realized experimentally either by determining the four-momentum vectors of 
the  outgoing
proton, neutron,  and $\eta^{\prime}$   or by measuring  the four-momentum
of the spectator proton. 
Since at present none of the experimental facilities installed at COSY can fulfill the first requirement
only the case with the registration of the spectator proton will be considered.
From the measurement of the momentum vector of the spectator
proton one can infer the momentum vector of the neutron at the time of the reaction, and hence
calculate the excess energy. 
The distribution 
of the excess energy in a quasi-free $pn\rightarrow pn\eta^{\prime}$ reaction
is presented in Figure~\ref{Q_and_mass_off}a. 
In the framework of the so-called impulse approximation
the spectator proton is a physical particle, hence  it must be on its mass shell.
This implies, however, that the reacting neutron is off its mass shell, and hence
the extrapolation from the quasi-free to the free $pn\rightarrow pn\eta^{\prime}$
must be done with 
care. The distribution of the off-shell mass
of the interacting neutron is shown in Figure~\ref{Q_and_mass_off}b. It can be seen
that the maximum of this spectrum differs only by about 3~MeV from the free neutron mass 
($m_{n}=939.57$~MeV), however on the average it is off by about 9~MeV.
 Measurements performed at the CELSIUS and TRIUMF accelerators 
 for the $pp\rightarrow pp\eta$~\cite{calenpdeta}
 and $pp\rightarrow d\pi^{+}$~\cite{duncan} reactions, respectively,
 have shown that within the 
 statistical errors there is no difference between the total cross section 
 of the  free and quasi-free  processes. The quasi-free production
 were realized utilizing a deuteron target.
This observation allows to anticipate that in the case of the planned
study of the $\eta^{\prime}$ in the $pd\rightarrow p_{sp}np\eta^{\prime}$ reaction,
the measured total cross section for the quasi-free
$pn\rightarrow pn\eta^{\prime}$ reaction will not differ  from the on-shell one.
In fact, the difference between off-shell and on-shell cross section 
of the $\eta^{\prime}$ meson production should be even smaller than 
in the case of the $\eta$ and $\pi$, since in the former case the total energy of the
interacting nucleons is much larger than in the latter one,
and the mean difference between the off-shell and on-shell neutron mass 
remains the same.

Other nuclear effects in case of the production on the neutron
bound in the nucleus are rather of minor importance.
The effect of the reduction of the beam flux on a neutron due to the presence
of the proton in the deuteron, referred to as a shadow effect, 
decreases the total cross section by about $4.5\%$
in case of 
$\eta$ production~\cite{chiavassa94}.
Thus it should also have a minor influence on the evaluation of the total cross
section for the $pn\rightarrow pn\eta^{\prime}$ reaction.
Similarly the reduction of the total cross section due to the reabsorbtion 
of the produced $\eta^{\prime}$ on the spectator proton should be
negligible due to the weak~\cite{swave} 
proton-$\eta^{\prime}$ interaction. Even in case of the $\eta$
meson, which interacts with protons much more strongly, 
this effect was found to be only about $3\%$~\cite{chiavassa94}.

In order to identify the production of the $\eta^{\prime}$ meson in a
proton-deuteron collision it is necessary either to measure the
decay products of the meson or to register the outgoing nucleons 
and nuclei. At present only the second possibility can be considered
at COSY experiments due to the lack of an appropriate 
detector\footnote{A photon detector is planned
         to be build and installed at the ANKE facility, 
         however, first experiments with
         this apparatus are foreseen for the end of the year 2004~\protect\cite{anke}.
        }.
The requirement to register  the spectator proton excludes the
possibility of performing such an experiment by means of any 
external facility which utilize liquid or solid targets.
\begin{figure}[H]
\hspace{0.18\textwidth}\epsfig{file=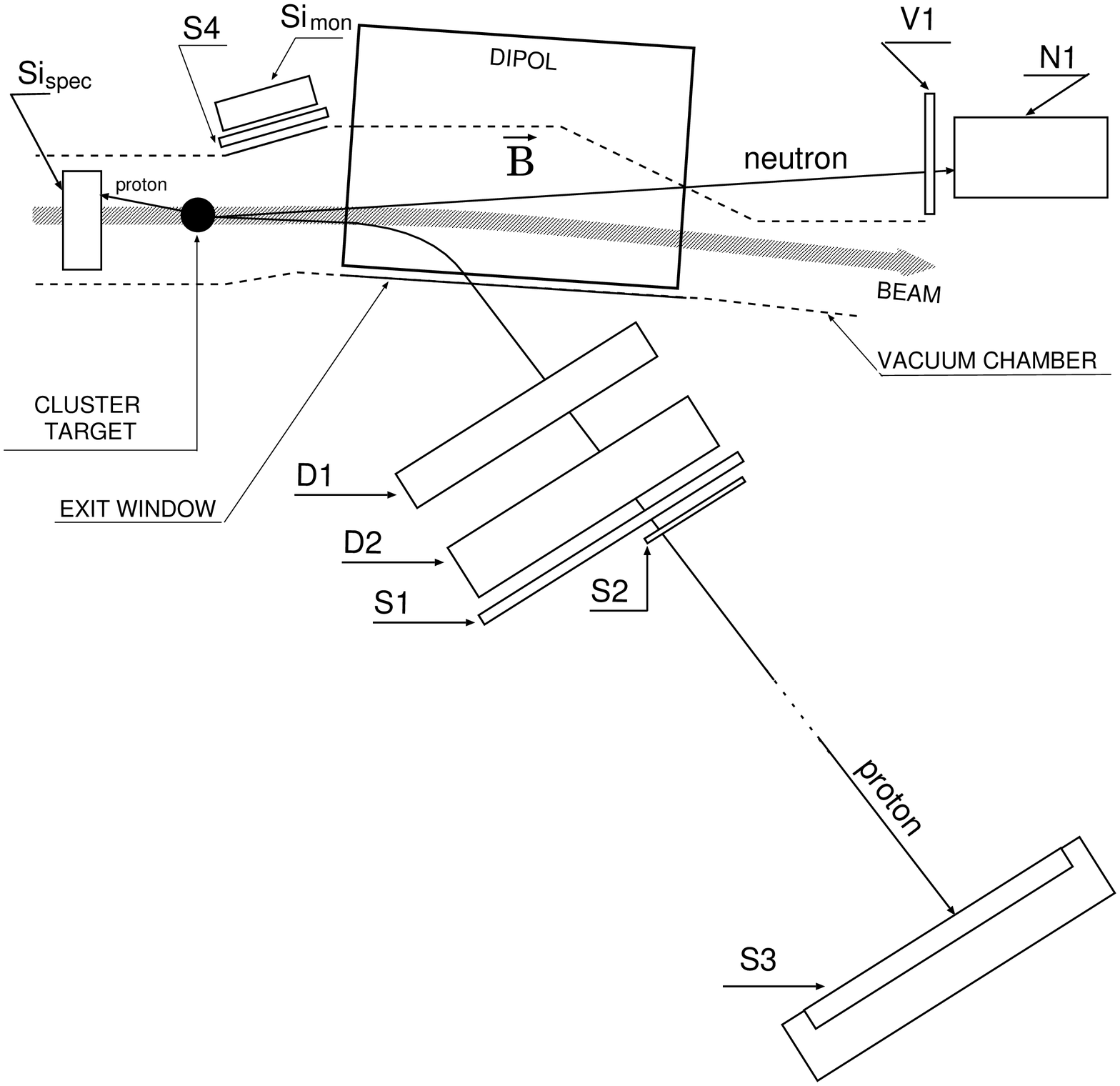,height=9cm,angle=0}
\vspace{-3.5cm}

\parbox{0.55\textwidth}{
            \caption{\small  Schematic view of the COSY-11 detection setup~\protect\cite{brau96}. 
             Only detectors needed for the measurements of the  reaction
             $pd\rightarrow p_{sp}pn\eta(\eta^{\prime})$
             are shown. \protect\\
             D1, D2 denote the drift chambers; S1, S2, S3, S4 and V1 the scintillation detectors;
             N1 the neutron detector and Si$_{mon}$ and Si$_{spec}$ silicon strip detectors
             to detect elastically scattered and spectator protons, respectively.
             \label{detectionsystem}
            }
          }
\end{figure}
At present, COSY-11 is the only internal facility at COSY, 
where the outgoing proton and neutron can be measured simultaneously.
Since for the proposed study  the ratio 
$R_{\eta^{\prime}}=\sigma(pn\rightarrow pn\eta^{\prime})/\sigma(pp\rightarrow pp\eta^{\prime})$
has to be determined it would be extremely advisable to perform 
both the production on protons and on neutrons by means of the same detection
system in order to minimize systematical uncertainties. 

During the last years close to threshold total cross sections for the 
$pp\rightarrow pp\eta^{\prime}$ reaction have been succesfully
measured at the COSY-11 facility~\cite{moskalpl,moskalprl}.
Thus the proposed investigation is planned to be performed using the COSY-11
detection system with an additional silicon strip detector for the registration of 
the spectator proton. The shape of the COSY-11 scattering chamber close to the
target allows for the installation of the spectator detectors~\cite{bilger}
 which were 
already used succesfully 
at the CELSIUS accelerator for tagging of quasi-free proton-neutron
interactions. 

In Figure~\ref{detectionsystem} a schematic view of the COSY-11 detection system
together with the spectator detector is presented. 
The spectator detector consists of  four modules with two layers of 0.3~mm
thick silicon strips. Each module is further divided into six parts 
each of them with 3 strips of 20~mm~x~5~mm.
The physical properties of the detector were implemented into the 
COSY-11 Monte-Carlo programme and detailed simulations of the 
$pd\rightarrow p_{sp}pn\eta^{\prime}$ reaction were performed, taking into account the 
momentum and dimension spread of the COSY beam~\cite{moskalnim},
as well as the dimensions
of the cluster target~\cite{domb97}
and the known resolution of the time and position 
measurements of the standard COSY-11 detectors.

\begin{figure}[h]
      \vspace{-0.0cm}
      \parbox{0.5\textwidth}{\epsfig{file=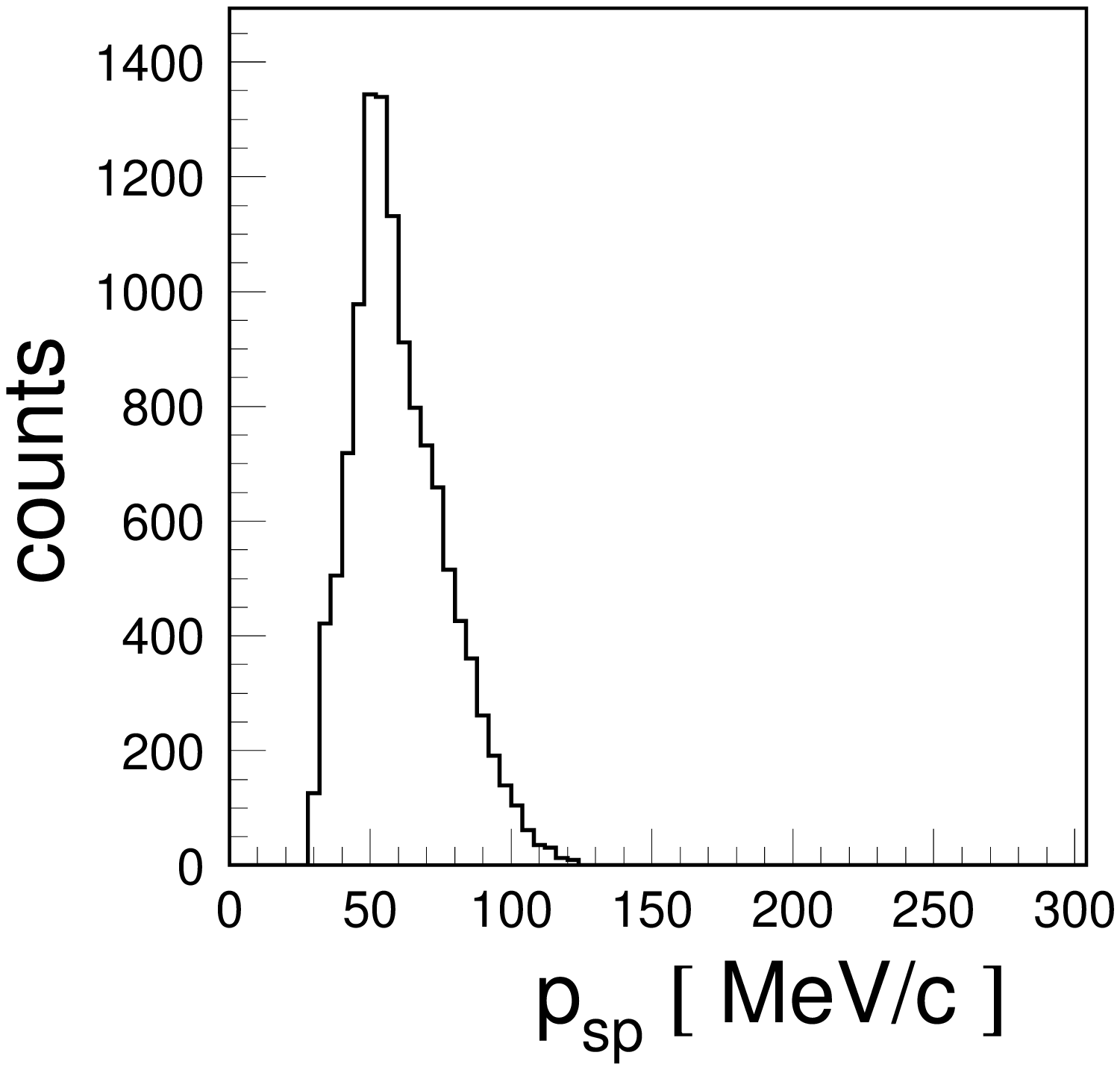,width=0.5\textwidth}}
      \parbox{0.5\textwidth}{\epsfig{file=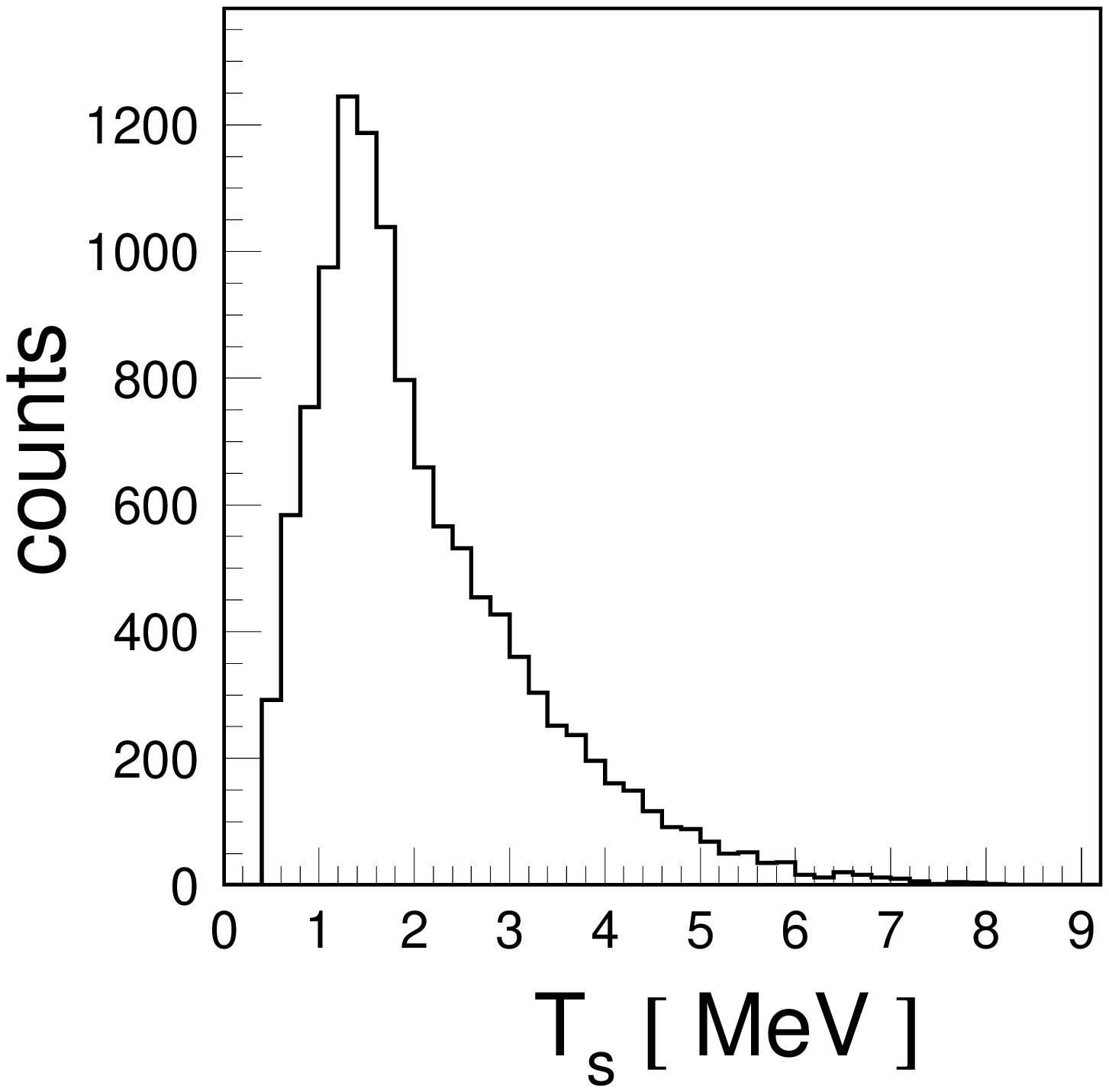,width=0.5\textwidth}}
       
       \vspace*{-0.9cm}
 
       \parbox{0.45\textwidth}{\mbox{} }\hfill
       \parbox{0.51\textwidth}{\large a) }\hfill
       \parbox{0.04\textwidth}{\large b) }

      \vspace*{-1.0cm}

      \parbox{0.5\textwidth}{\epsfig{file=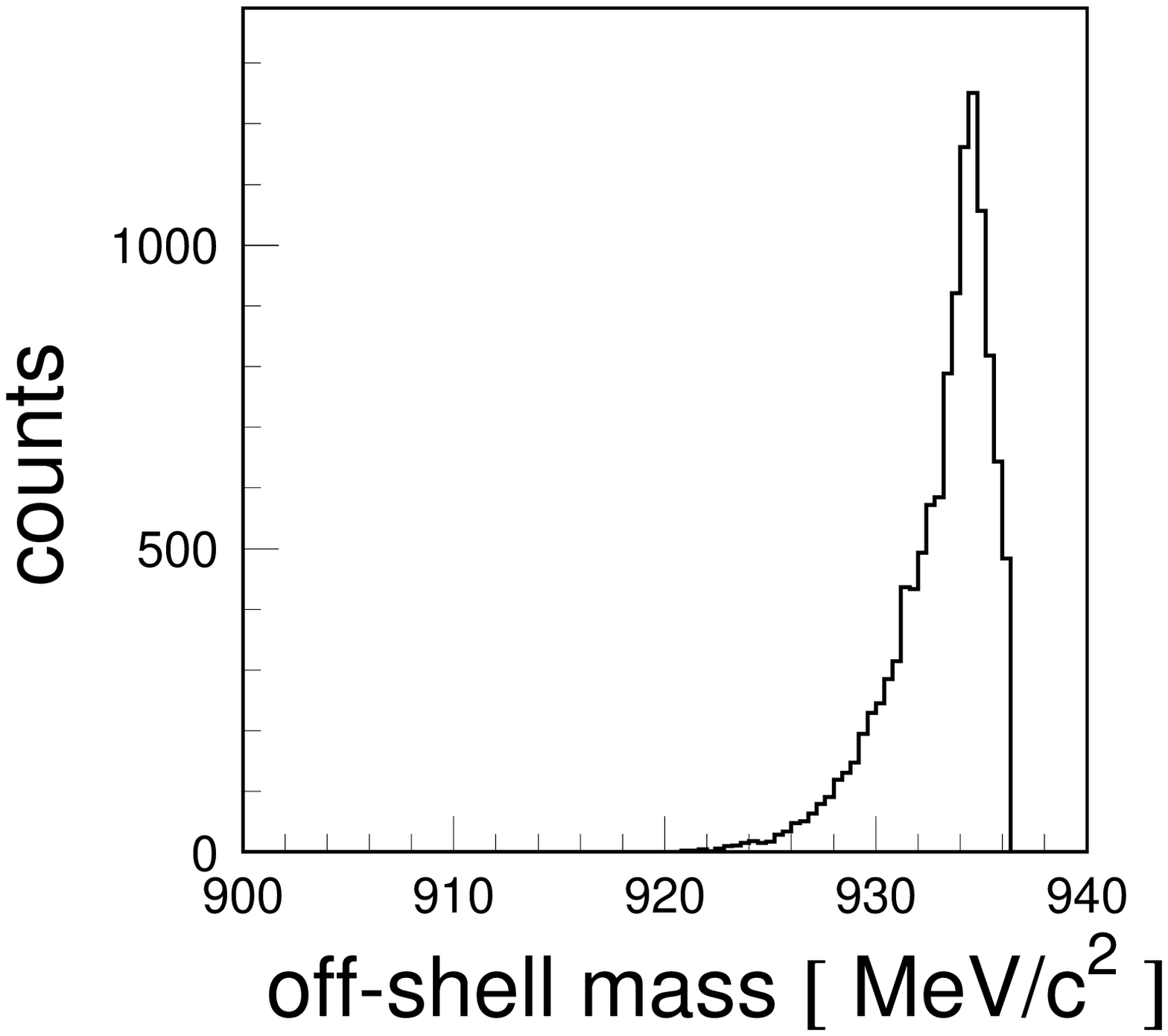,width=0.5\textwidth}}
      \parbox{0.5\textwidth}{\caption{ \small
            (a) Momentum and (b) kinetic energy  distributions
                 of the spectator protons 
                registered in coincidence with the forward scattered neutron and proton. \ \ \
            (c) Spectrum of the  off-shell mass of the interacting neutron,
                provided that all nucleons from the $pd\rightarrow p_{sp}pn\eta^{\prime}$
                reaction have been measured with the COSY-11 detection system.
            \label{spec_and_off_exp}
         }
         }
       
       \vspace*{-0.9cm}
 
       \parbox{0.45\textwidth}{\mbox{} }
       \parbox{0.05\textwidth}{\large c) }
\end{figure}

Figure~\ref{spec_and_off_exp}a depicts the spectrum of the spectator proton momentum
which will be registered in coincidence with the forward flying proton and neutron.
The fact that spectator protons with momenta larger than 130~MeV/c 
will not be registered 
has the advantage that events with neutron masses
differing much from its on-shell value will be omitted 
(compare figures~\ref{Q_and_mass_off}b and~\ref{spec_and_off_exp}c).
A $300~\mu$m thick silicon detector can absorb protons up to a momentum of about 106~MeV/c
corresponding to a kinetic energy  of T$_{s}\approx$~6~MeV.
Both layers together can absorb protons up to a momentum of 130~MeV/c (T$_{s}\approx 9$~MeV),
however, protons with momenta lower than 30~MeV/c (T$_{s}\approx$~0.5~MeV)
cannot be distinguished
from  noise. Thus this detector allows the registration and identification
of about 70$\%$ of the spectator protons hitting its sensitive 
area~(see Figure~\ref{impuls_and_spec_mom}).
On the other hand, Figure~\ref{spec_and_off_exp}b shows that the fraction 
of spectator protons having kinetic energy larger than 6~MeV and
registered simultaneously  with the forward scattered neutron and proton
is very small. Consequently, most of the spectator protons will be 
fully stopped already in the first layer. This gives the opportunity
to get rid of the background due to fast protons and pions crossing
both layers of the detector, just by considering in the off-line analysis
only those events which have no signal in the second layer.

\begin{figure}[h]
       \vspace{-1.0cm}
       \parbox{0.5\textwidth}{\epsfig{file=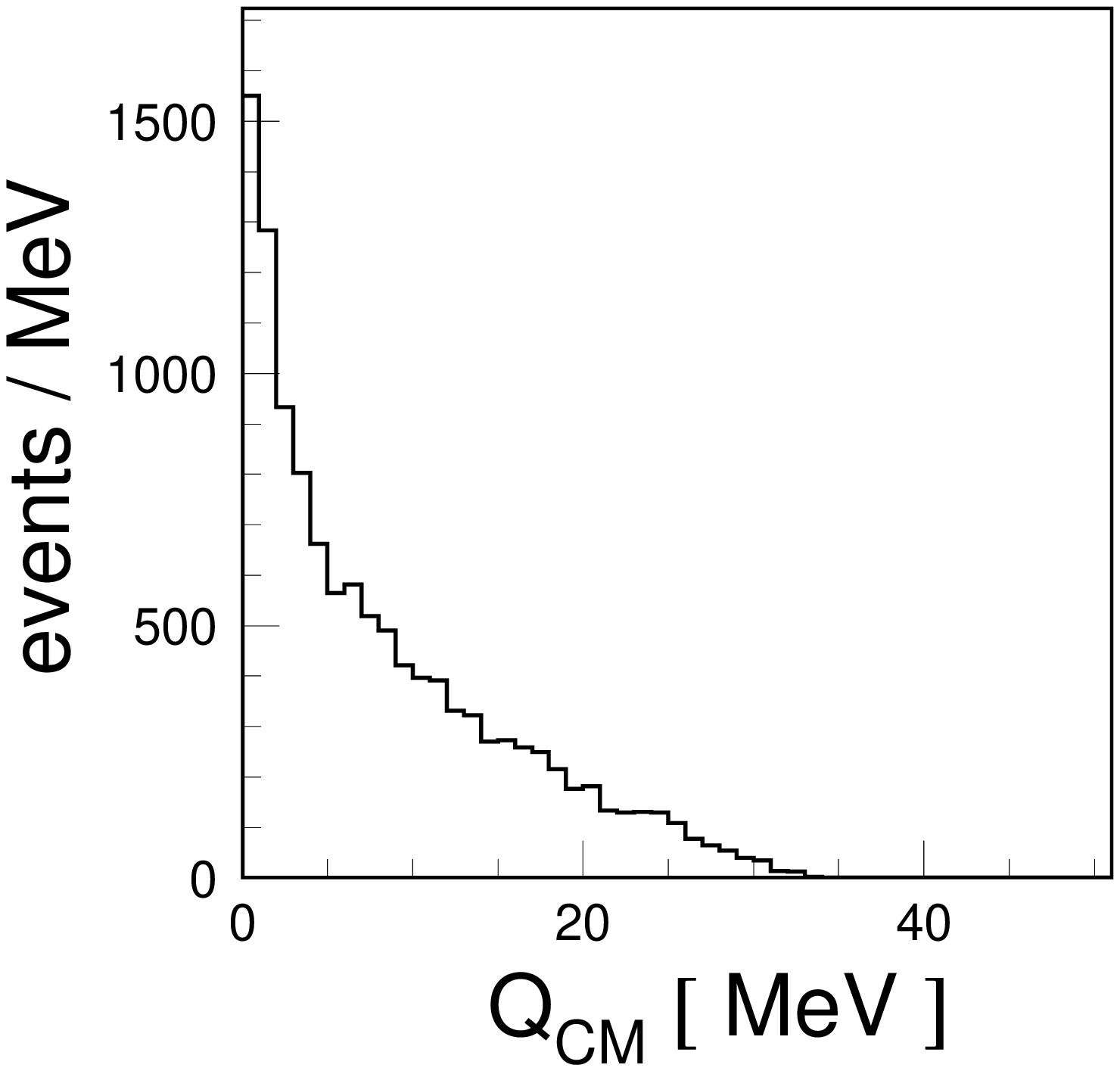,width=0.5\textwidth,angle=0}}
       \parbox{0.5\textwidth}{\epsfig{file=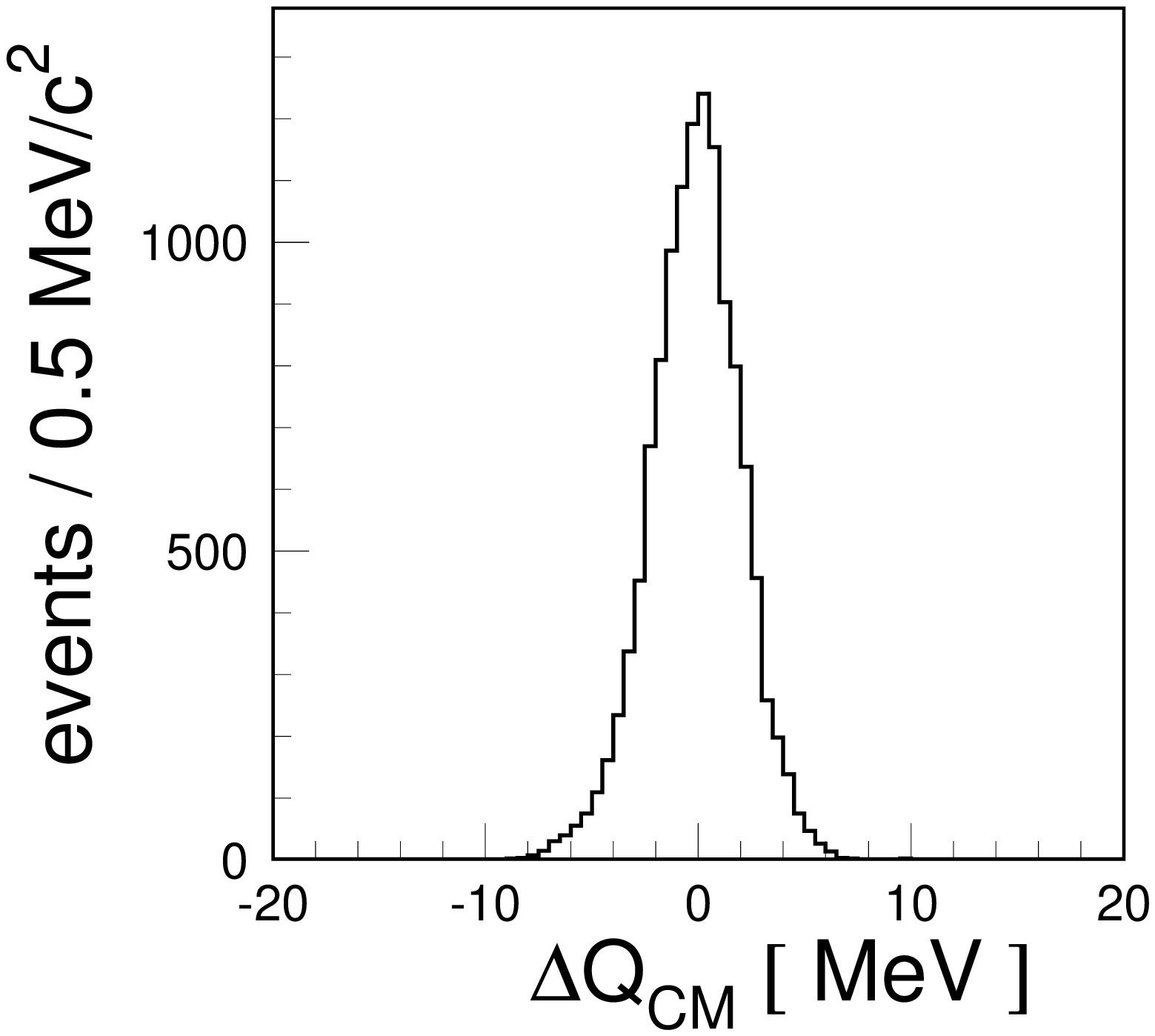,width=0.5\textwidth,angle=0}}

       \vspace*{-0.9cm}

       \parbox{0.45\textwidth}{\mbox{} }\hfill
       \parbox{0.51\textwidth}{\large a) }\hfill
       \parbox{0.04\textwidth}{\large b) }

   \caption{ \small
         (a) Distributions of the excess energy $Q_{CM}$
           for the quasi-free $pn\rightarrow pn\eta^{\prime}$
           reaction as would be measured with the COSY-11 detection system
           at a beam momentum of 3.350~GeV/c. The shape corresponds roughly
           to the energy dependence of the detection efficiency.
           The Figure shows the number of reconstructed  events per 1~MeV bin
           out of the $5 \cdot 10^{6}$ events generated in the target.
         (b) Difference between the generated and reconstructed excess energy.
             Calculations were performed assuming a target diameter of 0.9~cm~\protect\cite{domb97},
             and standard deviations of 0.2~cm and 0.4~cm for the horizontal and vertical
             beam spread, respectively. The applied values correspond to a realistic estimation of the
             beam parameters based on the data from previous COSY-11 experiments~\cite{moskalnim}.
             The spectator detector modules were positioned behind the target, five centimeters
             from the beam, as it is indicated in Figure~\ref{detectionsystem}.
             \label{Q_and_dQ_Qexp}
         }
\end{figure}

\begin{figure}[h]
\parbox{0.6\textwidth}{\centerline{\epsfig{file=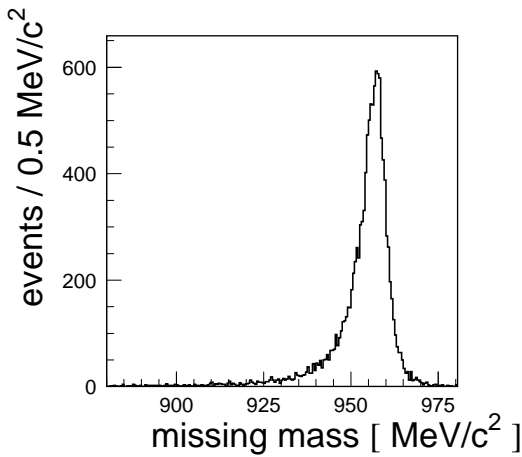,height=8.0cm,angle=0}}}
\parbox{0.4\textwidth}{\caption{ \small The missing mass distribution 
 with respect to the $pn$ subsystem from the reaction $pn\to pn \eta^{\prime}$
as reconstructed from the
generated events under the described assumptions of the 
          time and position resolution of the COSY-11 detectors.
          \label{missing_mass}
         }}
\end{figure}

Both the position of the spectator detector and the proton beam momentum were optimized
such that most of the registered events have an excess energy between 
0~MeV and 25~MeV. This is the range for which the total cross section 
of the $pp \rightarrow pp\eta^{\prime}$ has been studied at COSY-11~\cite{moskalpl}.
Increasing the beam momentum in general shifts the maximum 
of the excess energy distribution (fig.~\ref{Q_and_mass_off}a)
to larger values, but simultaneously installing the spectator detector 
upstream the target (see Figure~\ref{detectionsystem}), such that the spectator proton
 is moving opposite to
the proton beam, the excess energy is decreasing, since the neutron is escaping the beam proton.
Positioning of the tagger detector upstream the target also reduces drastically 
its irradiation.
The distribution of the excess energy for  events registered
under the chosen geometry is shown in Figure~\ref{Q_and_dQ_Qexp}a. In the first 
approximation it reflects the dependence 
of the COSY-11 acceptance
on the excess energy. 
 Figure~\ref{Q_and_dQ_Qexp}b
depicts the difference between the generated and reconstructed excess energy.               
The standard deviation of the obtained resolution amounts to 2~MeV, and is due to 
the finite dimension of the beam and target and  the granularity of the spectator detector.
In these simulations the distance between the beam  and the detector was 
adjusted to be 5~cm as an optimum regarding both the coverage of the solid angle and the
 angular resolution. However, the accuracy  of the excess energy  reconstruction
may be improved significantly if needed, since the scattering chamber allows
for the installation of spectator detectors at a distance of 10~cm from the 
beam.
Apart from the excess energy, in order to identify the reaction
it is necessary
to determine the momentum vectors of the registered protons and neutrons.
Proton momenta will be reconstructed by tracking back the proton trajectory to the target
point, and the neutron momenta will be determined from the time of flight
between the target and neutron detector and the angle defined by the middle of the
hit segment.  The granularity of the neutron detector allows to determine 
the horizontal position with an accuracy of $\pm 4.5$~cm. For the time resolution
of one segment (11 scintillation and 11 lead plates) a
conservative value of 0.5~ns (standard deviation) was assumed.
The missing mass spectrum, reconstructed from events for which signals from 
the spectator 
detector as well as the forward scattered protons and neutrons were registered, is shown
in Figure~\ref{missing_mass}. An obtained mass resolution amouts to 7.6~MeV (FWHM)
which is only 3 times the resolution of the measurements for 
the $pp\rightarrow pp\eta^{\prime}$ reaction at an excess energy of~23.6~MeV.
\vspace{-1.5cm}

\vspace{-1cm}

Assuming a luminosity of 3$\cdot 10^{ ~30}~$cm$^{-2}$s$^{-1}$,  and
taking into account i) the detection efficiency which decreases with excess
energy as shown in Figure~\ref{Q_and_dQ_Qexp}a and ii) the energy dependence of the
total cross section as determined for the $pp\rightarrow pp\eta^{\prime}$ reaction,
we calculate the number of quasi-free $pn \rightarrow pn \eta^{\prime}$ events
which will
be measured per day. The calculation can be performed modulo the absolute cross
section which in the one extreme scenario should be equal to the 
cross section of the $pp\rightarrow pp\eta^{\prime}$ reaction, and in the 
other extreme case should be enhanced by factor of about 6.5, as discussed in the
introduction.
The estimation results in 30 and 195 measured and reconstructed events per day
for the extreme scenarios. 
The energy dependence of the total cross section is predominantly determined
by the nucleon-nucleon final state interaction~\cite{nakayamaproc,kleefeld}, 
which approximately
has the same influence for $\eta$ and $\eta^{\prime}$ production.
Thus, the observation that the excitation function for the 
$pp\rightarrow pp\eta$ and $pn \rightarrow pn\eta$ 
reaction is approximately the same
justifies our assumption of the cross section dependence 
for the estimation of the counting rate.

A natural extension of the experiments with the close to threshold
production of $\eta$ and $\eta^{\prime}$ mesons in proton-proton
and proton-neutron collisions would be the creation of these mesons
in neutron-neutron reactions. This is a challenge for the future,
however, a first consideration concerning this possible study 
by utilizing a double quasi-free reaction has already been
described in reference~\cite{nn}.


\end{document}